\documentclass[aps,prd,a4paper,preprintnumbers,twocolumn,floatfix,showpacs,nofootinbib]{revtex4}
\usepackage{float,graphicx,multirow,bm,color,amsmath,amssymb,hyperref}
\hypersetup{
    colorlinks=true,
    linkcolor=green,
    citecolor=blue,
}

\begin{document}

\title{Generalizations of teleparallel gravity and local Lorentz symmetry}
\author{Thomas~P.~Sotiriou,$^1$ Baojiu Li$^{1,2}$ and John D.~Barrow$^1$}
\address{$^1$DAMTP, Centre for Mathematical Sciences, University of Cambridge, Cambridge CB3 0WA, UK\\ 
$^2$Kavli Institute for Cosmology Cambridge, Madingley Road, Cambridge CB3 0HA, UK}
\date{\today}

\begin{abstract}
We analyze the relation between teleparallelism and local Lorentz
invariance. We show that generic modifications of the teleparallel
equivalent to general relativity will not respect local Lorentz symmetry. We
clarify the reasons for this and explain why the situation is different in
general relativity. We give a prescription for constructing teleparallel
equivalents for known theories. We also explicitly consider a recently
proposed class of generalized teleparallel theories, called $f(T)$ theories of
gravity, and show why restoring local Lorentz symmetry in such theories
cannot lead to sensible dynamics, even if one gives up teleparallelism.
\end{abstract}

\maketitle

\address{$^1$DAMTP, Centre for Mathematical Sciences, University of Cambridge, Cambridge CB3 0WA, UK\\ 
$^2$Kavli Institute for Cosmology Cambridge, Madingley Road, Cambridge CB3 0HA, UK}




\section{Introduction}

\label{sect:Introduction}

A key feature of general relativity is observer independence, expressed
through general covariance. This makes it possible to formulate the theory
in terms of spacetime tensors, without making any reference to the tangent
space of each point of the spacetime manifold (even though this is where
vectors and tensors are naturally defined). Also, in general relativity, the
spacetime geometry can be fully described by the metric alone, because the
connection that defines parallel transport is assumed to be the Levi-Civita
connection of the metric. The effects of the gravitational interaction are
described in terms of the curvature of spacetime, which responds to the  
distribution and motion of mass-energy.

The avoidance of any reference to the tangent space in general
relativity is not mandatory. It is straightforward to introduce an
orthonormal basis for the tangent space at each point, the vierbein or
tetrad fields, $\mathbf{h}_{a}\left( x^{\mu }\right) $, project along this
basis and formulate the theory in terms of the projected quantities. This is
generally referred to as the tetrad or vierbein formalism. Such an approach
has advantages, especially when working with fermions. In the tangent-space
picture of general relativity, distances are measured with the flat metric,
but there still exists a non-vanishing connection with non-vanishing
curvature.

Alternatively, we could consider constructing a theory where, at least in a
suitable class of frames, the connection in the tangent space would have
zero curvature, without vanishing altogether.  This can be achieved if torsion is not zero and the corresponding connection is called the Weitzenbock
connection \cite{Weitzenbock}.
Such a theory is called teleparallel gravity \cite{Einstein,Aldrovandi}. The
simplest form of this theory is actually equivalent to general relativity 
\cite{Aldrovandi}. This appears surprising, given that the role of curvature
is so central in the latter. It is entirely consistent though, since the
zero-curvature Weitzenbock connection does not coincide with the Levi-Civita
connection of the metric. This will be explained in more detail below.

The teleparallel formulation of general relativity, which we will refer to
simply as teleparallel gravity below, allows a different physical
interpretation of the gravitational interaction in terms of torsion instead
of curvature. It has attracted interest in the past because it allows us to
interpret general relativity as a gauge theory.

Very recently, there have been proposals for constructing generalizations of
teleparallel gravity in Refs. \cite{Bengochea:2009, Yu:2010a,
Myrzakulov:2010a, Tsyba:2010, Linder:2010, Yi:2010b, Kazuharu:2010a,
Kazuharu:2010b, Myrzakulov:2010b, Yu:2010c, Karami:2010, Bengochea:2010, Yang:2010, Zheng:2010, Kazuharu:2010c, Dent:2010,ourpaper} which
followed the spirit of $f(R)$ gravity (see Ref.~\cite{Sotiriou:2008rp} for a
review) as a generalization of general relativity. That is, the lagrangians
of the theories were generalised to the form $f(T)$, where $f$ is some
suitably differentiable function and $T$ is the lagrangian of teleparallel
gravity. The interest in these theories was aroused by the claim that their
dynamics differ from those of general relativity but their equations are still second order in derivatives and, therefore, they might  be able to account for the accelerated expansion of the universe and remain free of pathologies. We showed in Ref.~\cite{us}, however, that this last
expectation was unfounded: these theories are not locally Lorentz invariant
and appear to harbour extra degrees of freedom.

Our aim here is to elaborate on the findings of Ref.~\cite{us}. More
specifically, we clarify below the role of violations of local Lorentz
invariance in generalized teleparallel theories. We provide an illustrative
example of an analogous situation with general covariance in ordinary field
theory. We will also explain why general relativity, which does respect
local Lorentz invariance, can nevertheless admit a teleparallel formulation.
It is argued that this is not a sign of the uniqueness for general
relativity, and a prescription is given for constructing teleparallel
equivalents of other known gravity theories. Finally, we focus on $f(T)$
theories and argue that, even if we decide to give up teleparallelism, such
actions would not make sense as descriptions of the dynamics of gravity if
local Lorentz symmetry was restored.

\section{Spacetime and tangent space descriptions}

\label{sec:form}

If we want to describe a spacetime in a coordinate basis, we need a metric $%
g_{\mu \nu }$ and a connection $\Gamma _{\phantom{a}\mu \nu }^{\lambda }$.
The connection does not have to be related to the metric. Instead of working
with a coordinate basis we could choose to associate a tangent space to each
spacetime point and work in terms of that tangent space. The vierbein or
tetrad fields, $\mathbf{h}_{a}\left( x^{\mu }\right) $, would then form an
orthonormal basis for the tangent space at each point of the manifold with
spacetime coordinates $x^{\mu }$. Latin indices label tangent space
coordinates while Greek indices label spacetime coordinates. All indices run
from $0$ to $3$. Clearly, $\mathbf{h}_{a}\left( x^{\mu }\right) $ is a
vector in the tangent space, and can be described in a coordinate basis by
its components $h_{a}^{\mu }$. So, $h_{a}^{\mu }$ also transforms as a
vector in spacetime. 

The spacetime metric, $g_{\mu \nu }$, is given by 
\begin{equation}  \label{eq:metric}
g_{\mu \nu }=\eta _{ab}h_{\mu }^{a}h_{\nu }^{b}\,,
\end{equation}%
where $\eta _{ab}=\mathrm{diag}(1,-1,-1,-1)$ is the Minkowski metric for the
tangent space. It follows that 
\begin{equation}
h_{a}^{\mu }h_{\nu }^{a}\ =\ \delta _{\nu }^{\mu },\ \ \ h_{a}^{\mu }h_{\mu
}^{b}\ =\ \delta _{a}^{b},
\end{equation}%
where Einstein's summation convention has been used.

A general connection cannot be described just in terms of the tetrad (in the
same way that the Christoffel symbols are not generically related to the
metric components). The following relations hold 
\begin{equation}
\Gamma _{\ \mu \nu }^{\lambda }\equiv h_{b}^{\lambda }\partial _{\nu }h_{\mu
}^{b}+h_{a}^{\lambda }A_{\phantom{a}b\nu }^{a}h_{\mu }^{b}\equiv
h_{b}^{\lambda }D_{\nu }h_{\mu }^{b}\,,
\end{equation}%
which also implicitly define the Lorentz covariant derivative $D_{\nu }.$
Here, $A_{\phantom{a}b\nu }^{a}$ is the spin connection and solving for it
we find 
\begin{equation}
A_{\phantom{a}b\nu }^{a}=h_{\lambda }^{a}\partial _{\nu }h_{b}^{\lambda
}+h_{\lambda }^{a}\Gamma _{\ \mu \nu }^{\lambda }h_{b}^{\mu }\equiv
h_{\lambda }^{a}\nabla _{\nu }h_{b}^{\lambda }\,,
\end{equation}%
where $\nabla _{\mu }$ denotes the covariant derivative associated with $%
\Gamma _{\ \mu \nu }^{\lambda }$. Note that $\Gamma _{\ \mu \nu }^{\lambda }$
is a Lorentz scalar (as long as $A_{\phantom{a}b\nu }^{a}$ is left
unrestricted). The torsion tensor is defined by 
\begin{equation}
T_{\phantom{a}\mu \nu }^{\lambda }\equiv \Gamma _{\ \nu \mu }^{\lambda
}-\Gamma _{\ \mu \nu }^{\lambda }\ .  \label{torten}
\end{equation}%
If we denote the Levi-Civita connection by 
\begin{equation}
\bar{\Gamma}_{\mu \nu }^{\lambda }\equiv \frac{1}{2}g^{\lambda \sigma
}\left( g_{\sigma \mu ,\nu }+g_{\sigma \nu ,\mu }-g_{\mu \nu ,\sigma
}\right) \,,
\end{equation}
then 
\begin{equation}
K_{\phantom{a}\mu \nu }^{\rho }\equiv \Gamma _{\ \mu \nu }^{\rho }-\bar{%
\Gamma}_{\mu \nu }^{\rho }\ 
\end{equation}%
is defined to be the contorsion tensor. Note that the Levi-Civita connection
does not have vanishing $A_{\phantom{a}b\nu }^{a}$; instead 
\begin{equation}
\bar{A}_{\phantom{a}b\nu }^{a}=h_{\lambda }^{a}\bar{\nabla}_{\nu
}h_{b}^{\lambda },
\end{equation}%
where a bar is used to denote all quantities associated with the Levi-Civita
connection. Finally, it is possible to show (after expressing the equation
in terms of the connection explicitly) that \emph{if} 
\begin{equation}
\nabla _{\lambda }g_{\mu \nu }=0,\   \label{metrcond}
\end{equation}%
then 
\begin{equation}
\label{torcontor}
K_{\phantom{a}\mu \nu }^{\rho }=\frac{1}{2}\left( T_{\mu \ \nu }^{\ \rho
}+T_{\nu \ \mu }^{\ \rho }-T_{\ \mu \nu }^{\rho }\right) .
\end{equation}%
That is, if the connection is metric compatible ({\em i.e.}~it has vanishing
non-metricity), then the contorsion tensor can be expressed in terms of the
torsion. From now on we will only consider metric compatible connections,
but not necessarily symmetric ones. Eq.~(\ref{torcontor}) can be solved for
the torsion to give 
\begin{equation}
T_{\rho \mu \nu }=K_{\rho \nu \mu }-K_{\rho \mu \nu }\,.  \label{contortor}
\end{equation}

Next, we define the tensor $S^{\rho \mu \nu }$ as 
\begin{equation}
S^{\rho \mu \nu }\equiv K^{\mu \nu \rho }-g^{\rho \nu }T_{\ \ \ \sigma
}^{\sigma \mu }+g^{\rho \mu }T_{\ \ \ \sigma }^{\sigma \nu },
\end{equation}%
and the associated invariant is 
\begin{eqnarray}  \label{Tdef}
T &\equiv &\frac{1}{2}S^{\rho \mu \nu }T_{\rho \mu \nu }=-S^{\rho \mu \nu
}K_{\rho \mu \nu }  \notag  \label{eq:T} \\
&=&\frac{1}{4}T^{\rho \mu \nu }T_{\rho \mu \nu }+\frac{1}{2}T^{\rho \mu \nu
}T_{\nu \mu \rho }-T_{\rho \mu }^{\ \ \rho }T_{\ \ \ \nu }^{\nu \mu }.
\end{eqnarray}%
Using the definitions given above, and without any restrictions on $A_{%
\phantom{a}b\nu }^{a}$ apart from those implied by metric compatibility for $%
\Gamma _{\phantom{a}\mu \nu }^{\lambda }$, we see that $T_{\phantom{a}\mu
\nu }^{\lambda }$ is a spacetime tensor and a Lorentz scalar, which makes $T$
both a spacetime scalar and a Lorentz scalar. On the other hand, 
\begin{equation}
R_{\phantom{a}\mu \lambda \nu }^{\rho }\equiv \partial _{\lambda }\Gamma _{%
\phantom{a}\mu \nu }^{\rho }-\partial _{\nu }\Gamma _{\phantom{a}\mu \lambda
}^{\rho }+\Gamma _{\phantom{a}\sigma \lambda }^{\rho }\Gamma _{\phantom{a}%
\mu \nu }^{\sigma }-\Gamma _{\phantom{a}\sigma \nu }^{\rho }\Gamma _{%
\phantom{a}\mu \lambda }^{\sigma },
\end{equation}%
and 
\begin{equation}
R_{\mu \nu }\equiv R_{\phantom{a}\mu \rho \nu }^{\rho }
\end{equation}%
are spacetime tensors and Lorentz scalars, and 
\begin{equation}
R\equiv g^{\mu \nu }R_{\mu \nu },
\end{equation}%
is both a spacetime scalar and a Lorentz scalar, just like $T$. The same
properties hold for $\bar{R}_{\phantom{a}\mu \lambda \nu }^{\rho }$, $\bar{R}%
_{\mu \nu }$ and $\bar{R}$. Using the definitions listed above, it is a
straightforward exercise to show that 
\begin{eqnarray}
\label{riemann}
R_{\phantom{a}\mu \lambda \nu }^{\rho }&=&\bar{R}_{\phantom{a}\mu \lambda \nu }^{\rho }+\bar{\nabla}_\lambda K^\rho_{\phantom{a}\mu\nu}-\bar{\nabla}_\nu K^\rho_{\phantom{a}\mu\lambda}\nonumber\\
&&+K^\rho_{\phantom{a}\sigma\lambda}K^\sigma_{\phantom{a}\mu\nu}-K^\rho_{\phantom{a}\sigma\nu}K^\sigma_{\phantom{a}\mu\lambda}\, ,\\
\label{riccitensor}
R_{\mu\nu}&=&\bar{R}_{\mu\nu }+\bar{\nabla}_\rho K^\rho_{\phantom{a}\mu\nu}-\bar{\nabla}_\nu K^\rho_{\phantom{a}\mu\rho}\nonumber\\
&& +K^\rho_{\phantom{a}\sigma\rho}K^\sigma_{\phantom{a}\mu\nu}-K^\rho_{\phantom{a}\sigma\nu}K^\sigma_{\phantom{a}\mu\rho}\, ,\\
R&=&\bar{R}+T+2\bar{\nabla} ^{\mu }\left( T_{\phantom{a}\mu \nu }^{\nu }\right) .\label{ricciscalar}
\end{eqnarray}%

Our aim in this section was just to give some basic definitions in two \emph{%
equivalent} descriptions of spacetime. So, the important message is that the
spacetime descriptions with a metric $g_{\mu\nu}$ and a (metric compatible,
but only due to our assumption) connection $\Gamma^\lambda_{\phantom{a}%
\mu\nu}$, is dual to a description which refers to a tangent space and uses
a tetrad $h_{a}^{\mu }$, and a spin connection $A^a_{\phantom{a}b\nu}$. We
intend to exploit the equivalence of the two descriptions in what comes next.

\section{Teleparallelism and local Lorentz invariance}

\label{sec:tele}

The main requirement of teleparallelism is that there exist a class of
frames where the spin connection vanishes, \emph{i.e.}~where 
\begin{equation}
A_{\phantom{a}b\nu }^{a}=0.
\end{equation}%
In these frames 
\begin{equation}
\Gamma _{\ \mu \nu }^{\lambda }\equiv h_{b}^{\lambda }\partial _{\nu }h_{\mu
}^{b}\ =\ -h_{\mu }^{b}\partial _{\nu }h_{b}^{\lambda }
\end{equation}%
which implies 
\begin{equation}
R_{\phantom{a}\mu \lambda \nu }^{\rho }=0,  \label{nocurv}
\end{equation}%
but nonzero torsion. A very important observation is that, since $R_{%
\phantom{a}\mu \lambda \nu }^{\rho }$ is a Lorentz scalar, if there exist
some class of frames where it is zero, then it will actually be zero in all
frames. But this requirement \emph{cannot} be imposed without introducing
some prior geometry.

Let us examine this in more detail. Suppose we are working in the tangent
space picture, where the fundamental fields are considered to be the tetrad $%
h_{a}^{\mu }$, and the spin connection $A_{\phantom{a}b\nu }^{a}$. Clearly,
if it is a characteristic of the theory that there exists a class of frames
in which $A_{\phantom{a}b\nu }^{a}=0$ then we can always choose to  work in
one of these frames (that is, define the theory as a preferred-frame theory 
\emph{a priori}). Then, we have $R=0$, and $T_{\phantom{a}\mu \nu }^{\lambda
}$ is manifestly not a Lorentz scalar anymore. The same holds for $T$. This
was the approach followed in Ref.~\cite{us} and many other papers in the
literature.

On the other hand, since $A_{\phantom{a}b\nu }^{a}$ is a connection, it can
be non-zero in other frames. Therefore, another option is to define all
quantities in a manifestly Lorentz covariant way, as done above, and enforce
the teleparallelism condition, \emph{i.e.}~that $A_{\phantom{a}b\nu }^{a}=0$
in some class of frames, as a constraint on the form of $A_{\phantom{a}b\nu
}^{a}$. Such constraints are usually imposed either by the explicit use of a
Lagrange multiplier, or implicitly by allowing only variations that respect
them when extremizing the action. Even though in this formulation  the
action can be made manifestly covariant, it is not really a way to restore
local Lorentz invariance at the level of the solutions due to the existence
of the constraint. This is best seen in the dual picture where the tangent
space is abandoned and the theory is described by the metric $g_{\mu \nu }$
and the connection $\Gamma _{\phantom{a}\mu \nu }^{\lambda }$ (clearly, the
Levi-Civita connection of the metric exists as well but it is not an
independent field). As mentioned earlier, in this picture, the requirement
of teleparallelism, that there be a class of frames where $A_{\phantom{a}%
b\nu }^{a}=0,$ translates to the requirement that $R_{\phantom{a}\mu \lambda
\nu }^{\rho }=0$ in \emph{all} frames because $R_{\phantom{a}\mu \lambda \nu
}^{\rho }$ is a Lorentz scalar. Suppose that the action of such a theory is
written in a manifestly (spacetime and of course Lorentz since we have
abandoned the use of tangent space) covariant way. Enforcing the constraint $%
R_{\phantom{a}\mu \lambda \nu }^{\rho }=0$ at the level of the field
equation (e.g. through a suitable covariant lagrange multiplier) implies the
existence of a second metric which is always forced to be flat. Obviously,
such a theory cannot generically respect local Lorentz covariance.

An illuminating example of an analogous situation in a much simpler theory
is that of a relativistic massless scalar field in flat spacetime, which is
usually described by the action 
\begin{equation}
S_{\phi }=\int d^{4}x\,\eta ^{\mu \nu }\partial _{\mu }\phi \partial _{\nu
}\phi \,.  \label{saction}
\end{equation}%
Variation with respect to $\phi $ yields 
\begin{equation}
\eta ^{\mu \nu }\partial _{\mu }\partial _{\nu }\phi =0\,.  \label{seos}
\end{equation}%
Now consider the action 
\begin{equation}
S_{\phi }^{c}=\int d^{4}x\sqrt{-g}\left( g^{\mu \nu }\bar{\nabla}_{\mu }\phi 
\bar{\nabla}_{\nu }\phi +M_{\rho }^{\phantom{a}\mu \lambda \nu }{\bar{R}}_{%
\phantom{a}\mu \lambda \nu }^{\rho }\right) \,.  \label{csaction}
\end{equation}%
Variation with respect to $\phi $ yields 
\begin{equation}
g^{\mu \nu }\bar{\nabla}_{\mu }\bar{\nabla}_{\nu }\phi =0\,,  \label{cseos}
\end{equation}%
whereas variation with respect to $A_{\rho }^{\phantom{a}\mu \lambda \nu }$
yields 
\begin{equation}
{\bar{R}}_{\phantom{a}\mu \lambda \nu }^{\rho }=0\,.  \label{flatg}
\end{equation}%
The last equation has the unique solution $g_{\mu \nu }=\eta _{\mu \nu }$.
Then, Eq.~(\ref{cseos}) becomes identical to Eq.~(\ref{seos}). There will
also be a third equation coming from the variation with respect to $g^{\mu
\nu }$, which will determine $M_{\rho }^{\phantom{a}\mu \lambda \nu }$.
However, the dynamics of $M_{\rho }^{\phantom{a}\mu \lambda \nu }$ become
irrelevant as there is no coupling to $\phi $, not even an indirect one
since Eq.~(\ref{flatg}) forces the metric to be flat.

The point of this example is to illustrate that we can write the action or
the field equations of a scalar field in flat space in a manifestly
covariant way. If action (\ref{csaction}) is taken at face value the theory
appears to be invariant under diffeomorphisms. However, at the level of the
solutions this theory is only invariant under global Lorentz
transformations, exactly like the theory described by action (\ref{saction}).

\section{Teleparallel formulation of general relativity and local Lorentz
invariance}

Let us now return to teleparallel gravity. Superficially, there appears to
be a contradiction in the statements made above: it was claimed in 
Sect.~\ref{sect:Introduction} that, on the one hand, teleparallel gravity can provide an
alternative formulation of general relativity, which is intrinsically a locally Lorentz
invariant theory. On the other hand, it was argued in the last section that
the requirement that there is a class of frames where the spin connection
vanishes, which is the cornerstone of teleparallelism, cannot generically be
enforced without violating local Lorentz invariance. However, there is no
real contradiction, and the resolution lies on the exact form of the
lagrangian (\ref{lagr}).

The action for (ordinary) teleparallel gravity is given by 
\begin{equation}  \label{lagr}
S_{T}\equiv \frac{1}{16\pi G}\int d^4x\,h\,T\ ,
\end{equation}%
in which $h=\sqrt{-g}$ is the determinant of $h_{a}^{\lambda }$ and $g$ is
the determinant of the metric $g_{\mu \nu }$, $G$ is the gravitational
constant. Eq.~(\ref{nocurv}) implies that $R=0$ in teleparallel theories,
and Eq.~(\ref{ricciscalar}) yields 
\begin{equation}  \label{TR}
T=-\bar{R}-2\bar{\nabla} ^{\mu }\left( T_{\phantom{a}\mu \nu }^{\nu }\right)
.
\end{equation}
Consequently, the action differs from the Einstein--Hilbert action only by a
boundary term, and will therefore lead to the same field equations. This is
why teleparallel gravity can be considered as an alternative formulation of
general relativity.

The presence of the boundary term is crucial. Without it, the action is the
Einstein--Hilbert action with the usual symmetries. Adding it and enforcing
the teleparallelism condition, the action becomes that of teleparallel
gravity, with lagrangian (\ref{lagr}), and is no longer locally Lorentz
covariant. In conclusion, the Einstein--Hilbert action, which of course
leads to a fully diffeomorphism invariant and Lorentz invariant theory, can
be written as the sum of two pieces: the teleparallel action and a boundary
term. Neither of these two pieces is locally Lorentz invariant once
teleparallelism is imposed (though they sum up to a locally Lorentz
invariant quantity). Imposing the last condition however, and formally
subtracting the boundary term, is crucial for the interpretation of the
theory as a teleparallel theory of gravity. Of course, this does not alter
at all the dynamical context of the theory, or its real symmetries in the
spacetime picture.

In conclusion, the reason that the teleparallel theory described by the
action (\ref{lagr}) does not really violate local Lorentz symmetry, even
though it respects teleparallelism, is because its action only differs from
a locally Lorentz covariant action by a boundary term. In fact, this action
turns out to be that of general relativity. Of course this cannot be a
property of a general teleparallel action. If a lagrangian is constructed
with the tetrad and the torsion tensor, then even if this lagrangian is a
spacetime and Lorentz scalar initially, it will not be a local Lorentz
scalar once teleparallelism has been imposed.

\section{Teleparallel formulation of gravity theories}

The previous discussion does not imply that general relativity is the only
gravity theory that can be cast into a teleparallel formulation and take on
a teleparallel interpretation. In fact, given eqs.~(\ref{riemann}), (\ref%
{riccitensor}) and (\ref{ricciscalar}) it should be clear that any action
constructed with curvature invariants of the metric can be cast into a
teleparallel formulation. Additionally, all teleparallel theories (\emph{i.e.}~theories whose
lagrangians are constructed with the curvature-free Weitzenbock connection
and the tetrad) whose action differs from a diffeomorphism invariant and locally
Lorentz invariant action only by a boundary term, will lead to locally
Lorentz invariant theories. This follows from straightforwardly generalizing the example of general
relativity considered above.

To understand this better, let us consider some simple examples. Let us
start from Brans-Dicke theory \cite{BD1961}, which is the best known alternative theory of
gravity. The action of the theory is
\begin{equation}
S_{BD}=\int d^{4}x\sqrt{-g}\left[ \phi \bar{R}-\frac{\omega _{0}}{\phi }\bar{%
\nabla}_{\mu }\phi \bar{\nabla}^{\mu }\phi \right] +S_{M}(g_{\mu \nu },\psi
),
\end{equation}%
where $\omega _{0}$ is the Brans-Dicke parameter, $S_{M}$ is the matter
action and $\psi $ collectively denotes the matter fields. Simply using Eq.~(%
\ref{ricciscalar}) and the fact that teleparallelism requires $R=0$, this
action can take the form of a teleparallel theory 
\begin{eqnarray}
S_{BD} &=&\int d^{4}x\,h\Big[-\phi T-\frac{\omega _{0}}{\phi }\nabla _{\mu
}\phi \nabla ^{\mu }\phi  \notag \\
&&\qquad \qquad +2\,T_{\phantom{a}\mu \nu }^{\nu }\nabla ^{\mu }\phi \Big]%
+S_{M}(g_{\mu \nu },\psi ),
\end{eqnarray}%
and can acquire a teleparallel interpretation. Note that we have discarded a
boundary term.

Another example is $f(\bar{R})$ gravity. The action for $f(\bar{R})$ gravity
is 
\begin{equation}
S_{f}=\int d^{4}x\sqrt{-g}f(\bar{R})+S_{M}(g_{\mu \nu },\psi ).
\label{faction}
\end{equation}%
As is well known \cite%
{Teyssandier:1983zz,Barrow:1988xi,Barrow:1988xh,Wands:1993uu}, as long as $%
f^{\prime \prime }(\bar{R})\neq 0$ this action can be brought into the form 
\begin{equation}
S_{f}=\int d^{4}x\sqrt{-g}\Big[\phi \bar{R}-V(\phi )\Big]+S_{M}(g_{\mu \nu
},\psi ),  \label{equivfR}
\end{equation}%
where 
\begin{equation}
V(\phi )=f(\chi )-\chi \phi ,
\end{equation}%
and $\chi $ implicitly defined through $\phi =f^{\prime }(\chi )$. The prime
denotes differentiation with respect to the argument. Again, using Eq.~(\ref%
{ricciscalar}) to replace $\bar{R}$, imposing teleparallelism and discarding
a boundary term, we get 
\begin{equation}
S_{f}=\!\!\int d^{4}x\sqrt{-g}\Big[-\phi T+2\,T_{\phantom{a}\mu \nu }^{\nu
}\nabla ^{\mu }\phi -V(\phi )\Big]+S_{M}(g_{\mu \nu },\psi ).
\label{equivRT}
\end{equation}

These simple examples demonstrate how we can construct teleparallel versions
of known gravity theories. On the other hand, we could also use $f(T)$
theories of gravity as a characteristic example of why generic \emph{ad hoc}
teleparallel actions will not respect local Lorentz invariance. Let us consider the
action 
\begin{equation}
S_{f(T)}\equiv \int d^{4}x\,h\,f(T)\ .  \label{flagr}
\end{equation}%
In an analogous manner to the procedure followed above for $f(\bar{R})$
gravity, this action can be brought into the form 
\begin{equation}
S_{f(T)}=\int d^{4}x\,h\Big[\phi T-V(\phi )\Big]+S_{M}(g_{\mu \nu },\psi ).
\label{equivfT}
\end{equation}%
To demonstrate this, first consider the action 
\begin{equation}
S_{1}=\int d^{4}x\,h\Big[f(\chi )-\phi (\chi -T)\Big]+S_{M}(g_{\mu \nu
},\psi ).  \label{S1}
\end{equation}%
Variation with respect to $\phi $ yields the algebraic constraint $\chi =T$.
Replacing this constraint in Eq.~(\ref{S1}) gives Eq.~(\ref{flagr}), implying the
dynamical equivalence of these two actions. On the other hand, variation
with respect to $\chi $ yields another algebraic constraint: $\phi
=f^{\prime }(\chi )$. Placing this constraint back in Eq.~(\ref{S1}), suitably
defining $V(\phi )$ and writing the action in terms of $\phi $ instead of $%
\chi ,$ yields the equivalent action Eq.~(\ref{equivfT}). Suppose now that we
want to use Eq.~(\ref{ricciscalar}) together with the teleparallelism
constraint $R=0$ in order to eliminate $T$ in favour of $\bar{R}$. We would
then get 
\begin{eqnarray}
S_{f(T)} &=&\int d^{4}x\sqrt{-g}\Big[-\phi \bar{R}+2\,T_{\phantom{a}\mu \nu
}^{\nu }\bar{\nabla}^{\mu }\phi   \notag  \label{equivfTR} \\
&&\qquad \qquad \qquad -V(\phi )\Big]+S_{M}(g_{\mu \nu },\psi ).
\end{eqnarray}%
It is the presence of the $T_{\phantom{a}\mu \nu }^{\nu }\bar{\nabla}^{\mu
}\phi $ term that leads to the violations of local Lorentz invariance as,
under the constraint of teleparallelism, $T_{\phantom{a}\mu \nu }^{\nu }$ is
not a Lorentz scalar anymore. Indeed, this term makes the difference between
actions Eq.~(\ref{equivfTR}) and Eq.~(\ref{equivfR}) or Eq.~(\ref{equivfT}) and Eq.~(\ref%
{equivRT}) (the sign differences are not important here as they can be
absorbed by a redefinition of $\phi $). Note that, action Eq.~(\ref{lagr})
differs from the Einstein-Hilbert action only by a boundary term, whereas
the difference between Eq.~(\ref{flagr}) and Eq.~(\ref{faction}) is not simply a
boundary term.

\section{A covariant version of $f(T)$ theories?}

We have established that general teleparallel theories will not respect
local Lorentz invariance and we have argued that this stems from the fact
that teleparallelism cannot generally be imposed without prior geometry. We
also saw that teleparallel $f(T)$ theories of gravity are typical examples
of theories that suffer from this problem. Suppose now that, for some
reason, we wish to restore local Lorentz invariance in these theories
without changing the form of the action. This can clearly be achieved only
by giving up teleparallelism. This is because, as explained in Sect.~\ref{sec:form}, 
if no restrictions related to teleparallelism are imposed
on $A_{\phantom{a}b\nu }^{a}$, then $T$ and consequently $f(T)$, will be local
Lorentz scalars.\footnote{%
One might wonder why imposing the teleparallelism constraint on $A_{%
\phantom{a}b\nu }^{a}$, which implies a constraint on $\Gamma _{\phantom{a}%
\mu \nu }^{\lambda }$, does not allow $T$ to be a Lorentz scalar, whereas
imposing that $\Gamma _{\phantom{a}\mu \nu }^{\lambda }$ does not contain
any part leading to nonmetricity does not cause such a problem. The reason
is that the latter restriction just expresses part of the connection in
terms of another dynamical field (the metric), whereas the former requires
the introduction of prior geometry as discussed in Sect.~\ref{sec:tele}.}

What would giving up the teleparallelism restriction on $A_{\phantom{a}b\nu
}^{a}$ mean for the dynamics for $f(T)$ theories? The best way to understand
the answer is to consider the duality between the $(h_{a}^{\mu },\,A_{%
\phantom{a}b\nu }^{a})$ and the $(g_{\mu \nu },\,\Gamma _{\phantom{a}\mu \nu
}^{\lambda })$ descriptions. The action is defined in the former as in Eq.~(%
\ref{flagr}), but since $h=\sqrt{-g}$ it can take the form 
\begin{equation}
S_{f(T)}\equiv \int d^{4}x\sqrt{-g}\,f(T)\ .  \label{flagr2}
\end{equation}%
What remains is to determine $T$ in terms of $g_{\mu \nu }$ and $\Gamma _{%
\phantom{a}\mu \nu }^{\lambda }$. Recall that $\Gamma _{\phantom{a}\mu \nu
}^{\lambda }$ is an independent connection which satisfies Eq.~(\ref%
{metrcond}). This implies that the part of this connection which is
independent of the metric is just the contorsion tensor $K_{\phantom{a}\mu
\nu }^{\rho }$. Therefore, the independent fields with respect to which we
must vary the action are $g_{\mu \nu }$ and $K_{\phantom{a}\mu \nu }^{\rho }$%
. Eq.~(\ref{contortor}) expresses the torsion in terms of the contorsion.
Taking a trace of the same equation yields 
\begin{equation}
T_{\phantom{a}\mu \rho }^{\rho }=-K_{\phantom{a}\mu \rho }^{\rho }\,.
\end{equation}%
Replacing the last expression in Eq.~(\ref{Tdef}), we can obtain an
expression for $T$ in terms of $g_{\mu \nu }$ and $K_{\phantom{a}\mu \nu
}^{\rho }$ only: 
\begin{equation}
T=-K^{\mu \nu \rho }K_{\rho \mu \nu }-K_{\phantom{ab}\sigma }^{\sigma \mu
}K_{\phantom{a}\mu \rho }^{\rho }\,.  \label{TK}
\end{equation}%
But with $T$ given in terms of $K_{\phantom{a}\mu \nu }^{\rho }$ by Eq.~(\ref%
{TK}), it becomes evident that the action (\ref{flagr2}) is dynamically
trivial, as it contains no derivatives of either of the two fundamental
fields: the metric and the contorsion. In fact, variations with respect to
the metric and the contorsion yield respectively 
\begin{eqnarray}
-f_{\phantom{a}\sigma (\mu }^{\prime \rho }K_{\phantom{a}\nu )\rho }^{\sigma
}-f_{\phantom{a}\mu \sigma }^{\prime \sigma }K_{\phantom{a}\nu \rho }^{\rho
}-\frac{f}{2}g_{\mu \nu } &=&S_{\mu \nu }\,,\quad   \label{f1} \\ \label{f2} 
(K_{\phantom{ab}\rho }^{\mu \nu }+K_{\phantom{a}\rho }^{\nu \phantom{a}\mu
}+K_{\phantom{ab}\sigma }^{\sigma \mu }\delta _{\rho }^{\nu }-K_{\phantom{a}%
\rho \sigma }^{\sigma }g^{\mu \nu })f^{\prime } &=&0\,,
\end{eqnarray}%
where we have taken into account the fact that the metric is symmetric, the
contorsion tensor is antisymmetric in the first two indices and 
\begin{equation}
S_{\mu \nu }\equiv -\frac{1}{\sqrt{-g}}\frac{\delta S_{M}}{\delta g^{\mu \nu
}}\,.
\end{equation}%
After some mathematical manipulations, and provided that $f^{\prime }(T)\neq 0
$ generally, Eq.~({\ref{f2}) yields 
\begin{equation}
K_{\phantom{a}\mu \nu }^{\rho }=0\,.
\end{equation}%
This implies that Eq.~(\ref{f1}) is trivially satisfied in vacuo and
inconsistent with the presence of matter. In addition, $g_{\mu \nu }$
remains indeterminate. The choice $f^{\prime }(T)=0$ clearly does not
improve things. }

These results should not come as a surprise and are not in conflict with the
fact that, after teleparallelism is enforced, the same action leads to
dynamical equations for the tetrad $h_{a}^{\mu }$. The reason for this is
that if $\Gamma _{\phantom{a}\mu \nu }^{\lambda }$ is to have zero curvature
then we must be able to express $K_{\phantom{a}\mu \nu }^{\rho }$ in terms
of the tetrad (much as the requirements of zero non-metricity and torsion
yield the expression for the Levi-Civita connection). This introduces
derivatives of the tetrad in the action and leads to a dynamical (yet
locally Lorentz-violating) theory.

Note that all of the arguments in this section could have be made in terms
of $T_{\phantom{a}\mu \nu }^{\rho }$ instead of $K_{\phantom{a}\mu \nu
}^{\rho }$, given the algebraic relation between the two quantities in Eq.~(%
\ref{torcontor}).

\section{Conclusions}

We have investigated the relation between teleparallelism and local Lorentz
symmetry violations, and have concluded that generically imposing
teleparallelism requires the introduction of prior geometry in the theory,
which leads to violations of local Lorentz symmetry. A special class of
teleparallel actions create an exception: those which differ from
diffeomorphism-invariant and locally-Lorentz-invariant actions (without
constraints) only by a boundary term. The Einstein-Hilbert action belongs
to this exceptional class. A prescription for constructing teleparallel
equivalents of known theories was also given. Finally, we focussed on $f(T)$
theories, which have attracted a lot of recent attention as possible dark
energy equivalents, and demonstrated that giving up teleparallelism in order
to restore local Lorentz invariance leads to a dynamically trivial theory,
which also becomes inconsistent if matter is added. Therefore, there seems
to be no way to get sensible dynamics from such an action, while
simultaneously satisfying local Lorentz invariance.

It would be interesting to explore further the generation of teleparallel
equivalents of known gravity theories and also to study their properties. It
could provide some insight into the interpretation of other gravity theories
as gauge theories.

\begin{acknowledgments}
We thank J.~G.~Pereira for helpful discussions. B.~Li is supported by
Queens' College, University of Cambridge and the Science and Technology
Facilities Council (\texttt{STFC}) rolling grant in DAMTP, Cambridge. 
T.~P.~Sotiriou is supported by a Marie Curie Fellowship.
\end{acknowledgments}

\end{document}